\documentclass[aps,prd,twocolumn,groupedaddress,amssymb,eqsecnum,showpacs,epsfig,nofootinbib]{revtex4}
\pdfoutput=1
\usepackage{graphicx}
\usepackage{bm}
\usepackage{dcolumn}
\usepackage{amsmath}
\usepackage{amsmath}
\usepackage{amssymb}

\numberwithin{equation}{section}
\usepackage{epsfig}
\def \lleq {\lower0.9ex\hbox{ $\buildrel < \over \sim$} ~}
\def \ggeq {\lower0.9ex\hbox{ $\buildrel > \over \sim$} ~}

\newcommand{\ben}{\begin{eqnarray}}
\newcommand{\een}{\end{eqnarray}}

\def \omm  {\Omega_{0 {\rm m}}}
\def \bomm  {{\bar \Omega}_{0 {\rm m}}}
\def \ommax  {\Omega_{0 {\rm m}\;max}}
\def \ommu  {\Omega_{0 {\rm m},u}}
\def \ommd  {\Omega_{0 {\rm m},d}}

\def \beq  {\begin{equation}}
\def \eeq  {\end{equation}}
\def \ber  {\begin{eqnarray}}
\def \eer  {\end{eqnarray}}

\newcommand {\ga} {\ {\raise-.5ex\hbox{$\buildrel>\over\sim$}}\ }
\newcommand {\la} {\ {\raise-.5ex\hbox{$\buildrel<\over\sim$}}\ }

\renewcommand{\(}{\left(}
\renewcommand{\)}{\right)}

\begin{document}
\newcommand{\newc}{\newcommand}
\newc{\be}{\begin{equation}}
\newc{\ee}{\end{equation}}
\newc{\ba}{\begin{eqnarray}}
\newc{\ea}{\end{eqnarray}}
\newc{\bea}{\begin{eqnarray*}}
\newc{\eea}{\end{eqnarray*}}
\newc{\D}{\partial}
\newc{\ie}{{\it i.e.} }
\newc{\eg}{{\it e.g.} }
\newc{\etc}{{\it etc.} }
\newc{\etal}{{\it et al.}}
\newc{\lcdm }{$\Lambda$CDM }
\newcommand{\nn}{\nonumber}
\newc{\ra}{\rightarrow}
\newc{\lra}{\leftrightarrow}
\newc{\lsim}{\buildrel{<}\over{\sim}}
\newc{\gsim}{\buildrel{>}\over{\sim}}
\title{Searching for a Cosmological Preferred Axis: \\Union2 Data Analysis and Comparison with Other Probes.}
\author{I. Antoniou and L. Perivolaropoulos}
\affiliation{Department of Physics, University of Ioannina,
Greece}
\date{\today}

\begin{abstract}
We review, compare and extend recent studies searching for evidence for a preferred cosmological axis. We start from the Union2 SnIa dataset and use the hemisphere comparison method to search for a preferred axis in the data. We find that the hemisphere of maximum accelerating expansion rate is in the direction $(l,b)=({309^\circ}^{+23^\circ}_{-3^\circ}, {18^\circ}^{+11^\circ}_{-10^\circ})$ ($\omm=0.19$) while the hemisphere of minimum acceleration is in the opposite direction $(l,b)=({129^\circ}^{+23^\circ}_{-3^\circ},{-18^\circ}^{+10^\circ}_{-11^\circ})$ ($\omm=0.30$). The level of anisotropy is described by the normalized difference of the best fit values of $\omm$ between the two hemispheres in the context of \lcdm fits. We find a maximum anisotropy level in the Union2 data of $\frac{\Delta \ommax}{\bomm}=0.43\pm 0.06$. Such a level does not necessarily correspond to statistically significant anisotropy because it is reproduced by about $30\%$ of simulated isotropic data mimicking the best fit Union2 dataset. However, when combined with the axes directions of other cosmological observations (bulk velocity flow axis, three axes of CMB low multipole moments and quasar optical polarization alignment axis), the statistical evidence for a cosmological anisotropy increases dramatically. We estimate the probability that the above independent six axes directions would be so close in the sky to be less than $1\%$. Thus either the relative coincidence of these six axes is a very large statistical fluctuation or there is an underlying physical or systematic reason that leads to their correlation.
\end{abstract}
\maketitle

\section{Introduction}
A large and diverse variety of cosmological observations during the past twenty years have established a standard cosmological model (\lcdm) which is based on the cosmological principle (homogeneity, isotropy, general relativity, cold dark matter + baryonic matter), flatness of space, the existence of a cosmological constant and gaussian scale invariant matter perturbations generated during inflation. This cosmological model makes clear and well defined predictions which have withstood the continuous and rapid improvement of cosmological observational tests. Prominent successes of the standard cosmological model include the following:
\begin{itemize}
\item
The Cosmic Microwave Background (CMB) angular power spectrum of perturbations \cite{Komatsu:2010fb} is overall in excellent agreement with the predictions of the standard model. However, a few issues related to the orientation and magnitude of low multipole moments (CMB anomalies) constitute remaining puzzles for the standard model \cite{Tegmark:2003ve,Copi:2010na,Bennett:2010jb,Schwarz:2004gk,Land:2005ad,Gruppuso:2010up,Sarkar:2010yj,Hanson:2009gu}.
\item
The statistics of the CMB temperature perturbation maps \cite{Smith:2009jr} are consistent with the prediction of gaussianity of the standard model.
\item
Observations of the recent accelerating expansion history of the universe \cite{Amanullah:2010vv} are consistent with the existence of a cosmological constant. Despite of the continuously improved data no need has appeared for more complicated models based on dynamical dark energy or modified gravity. The likelihood of the cosmological constant vs more complicated models has been continuously increasing during the past decade\cite{lpsniarev}.
\item
Observations of large scale structure are in good agreement with \lcdm \cite{Nesseris:2007pa} (basic statistics of galaxies \cite{TrujilloGomez:2010yh}, halo power spectrum \cite{Reid:2009xm}).
\end{itemize}
Despite of the above major successes the standard model is challenged by a few puzzling large scale cosmological observations \cite{Perivolaropoulos:2008ud} which may hint towards required modifications of the model. These challenges of \lcdm may be summarized as follows:
\begin{enumerate}
\item
{\bf Large Scale Velocity Flows:} \lcdm predicts significantly smaller amplitude and scale of flows than what observations indicate. It has been found that the dipole moment (bulk flow) of a combined  peculiar velocity sample extends \cite{Watkins:2008hf} on scales up to $100 h^{-1}Mpc$ ($z\leq 0.03$) with amplitude larger than $400 km/sec$. The direction of the flow has been found consistently to be approximately in the direction $l \simeq 282^\circ$, $b\simeq 6^\circ$. Other independent studies have also found large bulk velocity flows on similar directions \cite{Lavaux:2008th} on scales of about $100 h^{-1}Mpc$ or larger \cite{Kashlinsky:2008ut}. The expected $rms$ bulk flow in the context of \lcdm normalized with WMAP5 $(\omm,\sigma_8)=(0.258,0.796)$ on scales larger than $50h^{-1}Mpc$ is approximately $110 km/sec$. The probability that a flow of magnitude larger than $400km/sec$ is realized in the context of the above \lcdm normalization (on scales larger than $50h^{-1} Mpc$) is less than $1\%$. A possible connection of such large scale velocity flows and cosmic acceleration may be found in Ref. \cite{Tsagas:2009nh}.
\item
{\bf Alignment of low multipoles in the CMB angular power spectrum:} The normals to the octopole and quadrupole planes are aligned with
the direction of the cosmological dipole at a level inconsistent with Gaussian random, statistically isotropic skies at 99.7\% \cite{Copi:2010na}. The corresponding directions are: octopole plane normal $(l,b)=(308^\circ, 63^\circ)$ \cite{Bielewicz:2004en,Tegmark:2003ve}, quadrupole plane normal $(l,b)=(240^\circ, 63^\circ)$ \cite{Tegmark:2003ve,Frommert:2009qw}, CMB dipole moment $(l,b)=(264^\circ, 48^\circ)$ \cite{Lineweaver:1996xa}. A related effect has also been recently observed \cite{Kovetz:2010kv} by considering the temperature profile of 'rings' in the WMAP temperature fluctuation maps. It was found that there is a ring with anomalously low mean temperature fluctuation with axis in the direction $(l,b)=(276^\circ, -1^\circ)$ which is relatively close to the above directions (particularly that corresponding to the bulk velocity flows).
\item
{\bf Large scale alignment in the QSO optical polarization data:} Quasar polarization vectors are not randomly oriented over the sky with a probability often in excess of 99.9\%. The alignment effect seems to be prominent along a particular axis in the direction $(l,b)=(267^\circ, 69^\circ)$ \cite{Hutsemekers:2005iz}.
\item
{\bf Profiles of Cluster Haloes:} \lcdm predicts shallow low concentration and density profiles in contrast to observations which indicate denser high concentration cluster haloes \cite{Broadhurst:2004bi,Umetsu:2007pq}.
\item
{\bf Missing power on the low $l$ multipoles of the CMB angular power spectrum} which leads to a vanishing correlation function $C(\theta)$ on angular scales larger than $60^\circ$ \cite{Copi:2010na,Copi:2006tu,Sarkar:2010yj}
\end{enumerate}
In addition to the above large scale effects there are issues on galactic scales (missing satellites
problem \cite{Klypin:1999uc,Moore:2001fc,Madau:2008fr} and the cusp/core nature
of the central density profiles of dwarf galaxies
\cite{cuspygal}).
 In what follows we focus only on the kind of anomalies that appear to be related with the possible existence of a preferred axis. The rest are only mentioned for completeness but we can not exclude the possibility that a single mechanism could be responsible for seemingly unrelated anomalies. After all, before inflation, it would be hard to imagine that the same mechanism could be responsible for both the generation of primordial fluctuations on all scales and for the horizon problem (two seemingly unrelated issues).

Three of the above five large scale puzzles are large scale effects related to preferred cosmological directions (CMB multipole alignments, QSO polarization alignment and large scale bulk flows) which appear to be not far from each other. Their direction is approximately normal to the axis of the ecliptic poles $(l,b)=(96^\circ,30^\circ)$ and lies close to the ecliptic plane and the equinoxes. This coincidence has triggered investigations for possible systematic effects related to the CMB preferred axis but no significant such effects have been found \cite{Copi:2010na}.

Thus, unless there is a hidden common systematic \cite{Peiris:2010jd}, the existence of a cosmological preferred axis may be attributed to physical effects. An incomplete list of these effects is the following:
\begin{itemize}
\item
An anisotropic dark energy equation of state \cite{Zumalacarregui:2010wj,Koivisto:2005mm,Battye:2009ze} due perhaps to the existence of vector fields \cite{ArmendarizPicon:2004pm,EspositoFarese:2009aj}.
\item
Dark Energy and/or Dark matter perturbations on scales comparable to the horizon scale \cite{Rodrigues:2007ny,Jimenez:2008vs}. For example an off center observer in a 1Gpc void would experience the existence of a preferred cosmological axis through the Lematre-Tolman-Bondi metric \cite{Alexander:2007xx,GarciaBellido:2008nz,Biswas:2010xm,Dunsby:2010ts,Garfinkle:2009uf}.
\item
Deviations from the isotropic cosmic expansion rate induced by a fundamental violation of the cosmological principle eg through a multiply connected non-trivial cosmic topology \cite{Luminet:2008ew,Bielewicz:2008ga}, rotating universe coupled to an anisotropic scalar field \cite{Carneiro:2001fz}, non-commutative geometry \cite{Akofor:2007fv} or simply a fundamental anisotropic curvature\cite{Koivisto:2010dr}.
\item
Statistically anisotropic primordial perturbations\cite{ArmendarizPicon:2007nr,Pullen:2007tu,Ackerman:2007nb,ValenzuelaToledo:2010cs}. For example, inflationary perturbations induced by vector fields \cite{Dimopoulos:2008yv,Yokoyama:2008xw,Golovnev:2009ks,Bartolo:2009pa}. Note however that inflationary models with vector fields usually suffer from instabilities due to the existence of ghosts. \cite{ghosts}
\item
The existence of a large scale primordial magnetic field\cite{Kahniashvili:2008sh,Barrow:1997mj,Campanelli:2009tk}. Evidence for such a magnetic field has recently been found in CMB maps \cite{Kim:2009gi}.
\end{itemize}

In view of the convergence of the directions of the preferred axes mentioned above, it is important to investigate other cosmological datasets for hints of existence of preferred axes. Type Ia supernovae used as standard candles to investigate the recent expansion history of the universe constitute one such class of datasets.

Most previous studies searching for anisotropies in SnIa datasets have found no statistically significant evidence for anisotropies \cite{Blomqvist:2009ps,Blomqvist:2010ky,Tomita:2001gh,Cooke:2009ws,Kolatt:2000yg,Gupta:2007pb}. The latest such study \cite{Blomqvist:2010ky} used the latest and largest dataset available (Union2 \cite{Amanullah:2010vv} consisting of $557$ SnIa), to derive the angular covariance function of the standard candle magnitude fluctuations searching for angular scales where the covariance function deviates from 0 in a statistically significant manner. No such angular scale was found. This is a useful and ambitious approach which aims at identifying not only the existence of a possible anisotropy but also its detailed angular scale dependence despite of relatively small number of data in most angular scales considered.

An alternative approach is found in \cite{Schwarz:2007wf} where a statistically significant preferred axis was found using the hemisphere comparison method. This method, adopted also in the present study (see next section), amounts to fitting the \lcdm parametrization for $\omm$ on several pairs of opposite hemispheres and comparing the maximally asymmetric pair with the corresponding maximally asymmetric pair of isotropized similar datasets. The advantage of this method is that it optimizes the statistics since there is a large number of SnIa in each hemisphere. This is achieved at the cost of loosing all information about the detailed structure of the anisotropy.

The method was applied to four SnIa datasets \cite{Schwarz:2007wf}. The most prominent axis of maximal hemispheric asymmetry $(l,b)=(123^\circ,27^\circ)$ was found to be  close to the axis of the equatorial poles $(l,b)=(96^\circ,30^\circ)$. This alignment  between maximally asymmetric hemispheric Hubble diagrams and the equatorial frame was attributed to a systematic error by the authors of Ref. \cite{Schwarz:2007wf}. A second direction of maximum asymmetry $(l,b)=(235^\circ,15^\circ)$ in one of the datasets (the Gold04 \cite{Riess:2004nr}) considered in Ref. \cite{Schwarz:2007wf} was closer to the preferred axes of other observations discussed above but it was also discarded because the maximum asymmetry between the hemispheres coincided with the maximal asymmetry in the number of degrees of freedom.

In this study we apply a variation of the hemisphere comparison method to the Union2 dataset. Our goal is to identify the direction of the axis of maximal asymmetry for the Union2 dataset \cite{Amanullah:2010vv} (directions to the SnIa provided in Ref. \cite{Blomqvist:2010ky}) and evaluate its statistical significance by comparing with a large number of similar Monte Carlo isotropized datasets. We also compare the obtained direction of maximal asymmetry axis with the directions of preferred axes obtained with the other cosmological observations discussed above. In particular we find the probability that these axes would have the observed angular separation if they were uncorrelated.

The structure of this paper is the following: In the next section we describe in more detail the hemisphere comparison method and we apply it to the Union2 dataset. In section 3 we compare the direction of the axis of maximal asymmetry with the directions of  preferred axes from the other observations discussed above. Finally in section 4 we conclude summarize and discuss future prospects of this work.

\section{Maximum Asymmetry Axis of the Union2 Dataset}

The Union2 SnIa dataset \cite{Amanullah:2010vv} is a complilation consisting of 557 SNe Ia covering the redshift range
$z = [0.015, 1.4]$. It extends the Union dataset \cite{Kowalski:2008ez} by adding new SnIa data at low
and intermediate redshifts discovered by the CfA3 \cite{Hicken:2009dk} and SDSS-II Supernova Search \cite{Holtzman:2008zz},
respectively. It also includes six new SnIa discovered by the Hubble Space Telescope
at high $z$. Here we use the directions to the SnIa provided in Ref. \cite{Blomqvist:2010ky}. The angular distribution of the Union 2 dataset in galactic coordinates, is shown in Figure 1. The color of each point provides information about the redshift.

\begin{figure}[!b]
\begin{center}
\includegraphics[width=3.5in]{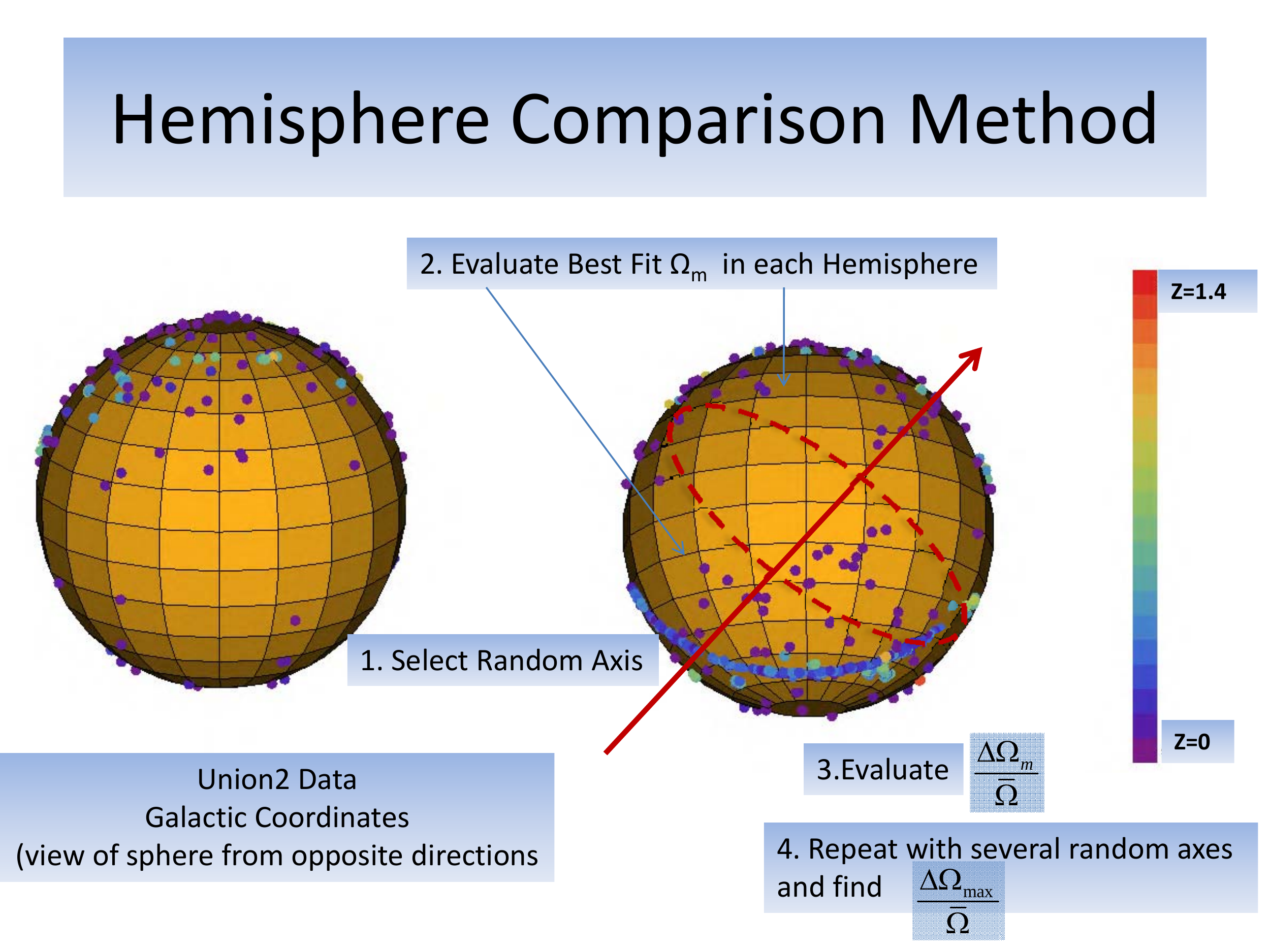}
\end{center}
\caption{\small The Union2 data in galactic coordinates and the Hemisphere Comparison Method. The color of each datapoint provides information about its redshift according to the legend on the right. Two opposite hemispheres are shown. The viewpoints are along the axis of maximum asymmetry discussed below.}
  \label{fig1}
\end{figure}

The Union2 data along with directions as presented in Ref. \cite{Blomqvist:2010ky} include the SnIa name, the redshift in the CMB rest frame, the distance modulus and its uncertainties (which include both the observational and the intrinsic magnitude scatter). They also include the equatorial coordinates (right ascension and declination) of each SnIa. It is straightforward to convert these coordinates to galactic coordinates or to usual spherical coordinates $(\theta,\phi)$ in the equatorial or galactic systems \cite{astrocalc}.

This dataset may be analyzed in the usual manner by applying the maximum likelihood method. The apparent magnitude
$m(z)$ is related to the Hubble-free luminosity distance $D_L(z)$ through \be
m_{th}(z,\omm)={\bar M} (M,H_0) + 5 log_{10} (D_L (z)) \label{mdl} \ee
where in a flat cosmological model \be D_L (z)= (1+z) \int_0^z
dz'\frac{H_0}{H(z';\omm)} \label{dlth1} \ee is the Hubble-free luminosity distance assumed here to be parameterized by \lcdm \ie
\be H(z)^2 =
H_0^2 [\omm (1+z)^3 + (1-\omm)] \ee
Also ${\bar M}$ is the magnitude zero
point offset which depends on the absolute magnitude $M$ and on the
present Hubble parameter $H_0$ as \ba
{\bar M} &=& M + 5 log_{10}(\frac{c\; H_0^{-1}}{Mpc}) + 25= \nn \\
&=& M-5log_{10}h+42.38 \label{barm} \ea The parameter $M$ is the
absolute magnitude which is assumed to be constant.

The data points of the Union2 dataset are given in terms of the distance
modulus \be \mu_{obs}(z_i)\equiv m_{obs}(z_i) - M \label{mug}\ee
The theoretical model parameter ($\omm$) is determined by minimizing  \be \chi^2 (\omm,\mu_0)= \sum_{i=1}^N
\frac{(\mu_{obs}(z_i) - \mu_{th}(z_i,\omm,\mu_0))^2}{\sigma_{\mu \; i}^2} \label{chi2} \ee where $\sigma_{\mu \; i}^2$
are the distance modulus uncertainties which include both the observational and the intrinsic magnitude scatter.
These uncertainties are
assumed to be gaussian and uncorrelated (we assume a diagonal covariance matrix and ignore systematics). The theoretical distance
modulus is defined as \be \mu_{th}(z_i,\omm,\mu_0)\equiv m_{th}(z_i,\omm) - M =5
log_{10} (D_L (z)) +\mu_0 \label{mth} \ee where \be \mu_0= 42.38 -
5 log_{10}h \label{mu0}\ee

As a test we have checked that our full sky analysis reproduces the results of Ref. \cite{Amanullah:2010vv} for the cases of no systematics. For example we obtain a full sky best fit value $\omm=0.27$ and we reproduced the contour Fig. 10a of Ref. \cite{Amanullah:2010vv}.
\begin{figure}[!b]
\begin{center}
\includegraphics[width=3.5in]{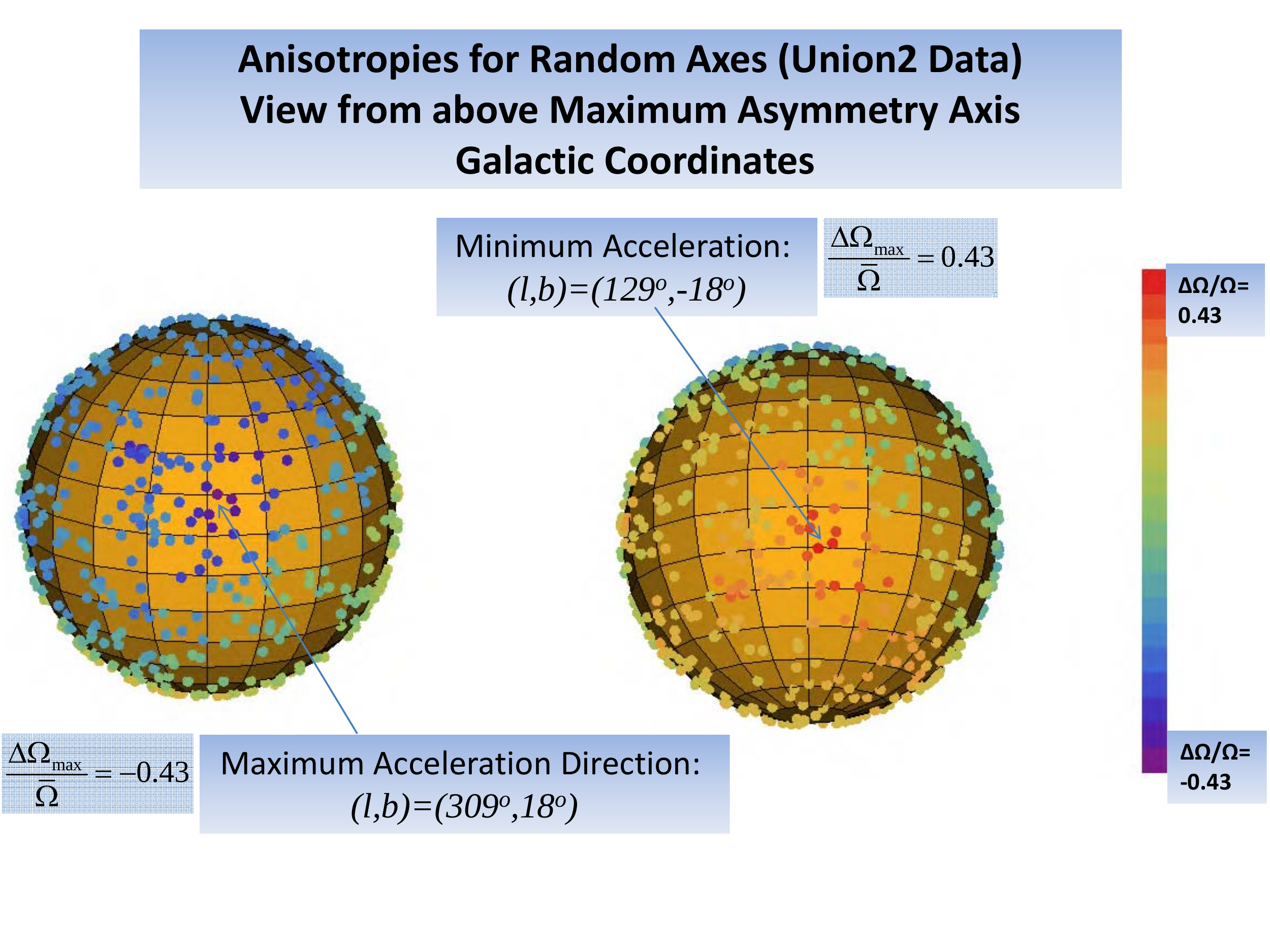}
\end{center}
\caption{\small The directions of the random axes considered are shown as dots on the unit sphere colored according to the sign and magnitude of the anisotropy level $\(\frac{\Delta \omm}{\bomm}\)^{U2}$. The hemisphere shown on the left (right) is the one corresponding to maximum (minimum) acceleration. The corresponding best fit values of $\omm$ are $\omm=0.19$ ($(l,b)=(309^\circ,18^\circ)$)  and $\omm=0.30$ ($(l,b)=(129^\circ,-18^\circ)$).}
  \label{fig2}
\end{figure}

The hemisphere comparison method implemented in our study involves the following steps (see also Fig. 1):
\begin{enumerate}
\item
Generate a random direction
\be
{\hat r}_{rnd}=(\cos\phi \; \sqrt{1-u^2},\sin\phi \; \sqrt{1-u^2},u) \label{randdir} \ee where $\phi\in [0,2\pi)$ and $u\in [-1,1]$ are random numbers with uniform probability distribution.
\item
Split the dataset under consideration into two subsets according to the sign of the inner product ${\hat r}_{rnd} \cdot {\hat r}_{dat}$ where ${\hat r}_{dat}$ is a unit vector describing the direction of each SnIa in the dataset. Thus one subset corresponds to the hemisphere in the direction of the random vector (defined as 'up') while the other subset corresponds to the opposite hemisphere (defined as 'down').
\item
Find the best fit values on $\omm$ in each hemisphere ($\ommu$ and $\ommd$). Use these values to obtain the anisotropy level quantified through the normalized difference \be \frac{\Delta \omm}{\bomm} \equiv 2\frac{\ommu-\ommd}{\ommu+\ommd} \label{dom} \ee
\item
Repeat for 400 random directions ${\hat r}_{rnd}$ and find the maximum standardized difference for the Union2 data $\(\frac{\Delta \ommax}{\bomm}\)^{U2}$. We also obtain the corresponding direction of maximum anisotropy.
\end{enumerate}
\begin{figure}[!b]
\begin{center}
\includegraphics[width=3.5in]{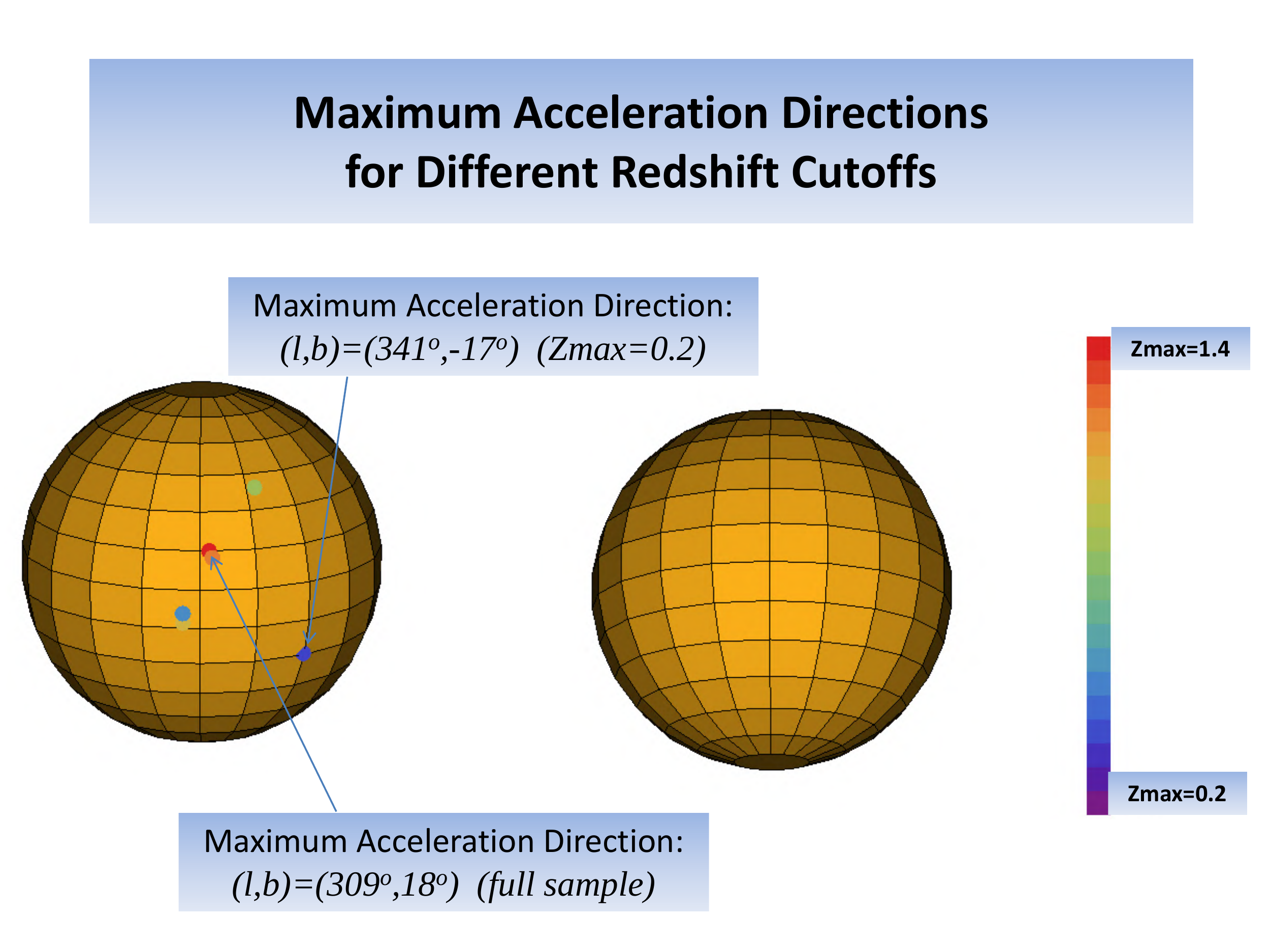}
\end{center}
\caption{\small The directions of maximum acceleration corresponding to the redshift cutoffs of Table 1.}
  \label{fig2}
\end{figure}
 Instead of the above algorithm involving random directions of the axes, we could have implemented a
uniform coverage of the sphere, utilizing equal area pixels
along the lines of Healpix \cite{Gorski:2004by}. In that case each axis would have a
fixed direction in the center of each equal area pixel.
However, we find the use of axes with random directions simpler
to implement in practice without any specific disadvantage
compared to the Healpix approach.

In order to maximize efficiency, the number of axes should be
approximately equal to the number of data points (SnIa) on each
hemisphere. The reason for this is that changing the direction
of an axis, does not change the corresponding $\frac{\Delta
\Omega_{0m,max}}{\bar \Omega_{0m}}$ until a data point is
crossed by the corresponding equator line. Such a crossing is
expected to occur when the direction of an axis changes by
approximately the mean angular separation between data points.
Thus, using more axes than the number of data points in a
hemisphere does not improve the accuracy of the determination
of the maximum anisotropy direction. Given that the number of
datapoints per hemisphere for the Union2 dataset is about 280
it becomes clear that using 280 test axes is close to the
optimal number of axes to use.  We have used used 400 axes in our analysis, well above the value of 280.

In order to derive the $1\sigma$
error in the maximum anisotropy direction, we first obtain the $1\sigma$ error $\sigma_{\Delta
\Omega}$ associated with $\frac{\Delta \Omega_{0m,max}}{\Omega_m}$.
This is of the form
\be
\sigma_{\Delta \Omega}=\frac{\sqrt{\sigma_{\Omega_{0m,u}}^2 + {\sigma_{\Omega_{0m,d}}^2}}}{\Omega_{0m,u}+\Omega_{0m,d}}=0.06 \label{errdom}
\ee
Notice that this is the error due to the uncertainties of the
supernova distance moduli propagated to the best fit $\Omega_{0m}$ on each
hemisphere and thus to $\frac{\Delta \Omega_{0m,max}}{\Omega_m}$. We
then identify all the test axes that correspond to an
anisotropy level within $1\sigma$ from the maximum anisotropy
level ie $\frac{\Delta \Omega_{0m}}{\bar
\Omega_{0m}}=\frac{\Delta \Omega_{0m,max}}{\bar \Omega_{0m}}\pm
\sigma_{\Delta \Omega}$. These axes apparently cover an angular
region corresponding to the $1\sigma$ range of the maximum
anisotropy direction.  Using a run with 400 test axes we find
the $1\sigma$ angular region for the hemisphere of maximum
acceleration is in the direction $l={309^\circ}^{+23^\circ}_{-3^\circ}$,
$b={18^\circ}^{+11^\circ}_{-10^\circ}$ (best fit $\omm=0.19$) while the hemisphere with the minimum acceleration is in the opposite direction $(l,b)=({129^\circ}^{+3^\circ}_{-23^\circ},{-18^\circ}^{+10^\circ}_{-11^\circ})$ (best fit $\omm=0.30$). The corresponding maximum anisotropy level is
\be \(\frac{\Delta \ommax}{\bomm}\)^{U2}=0.43 \pm 0.06 \label{maxlevel} \ee
Our results are shown in Fig. 2 where we present the directions of the random axes considered, as dots on the unit sphere colored according to the sign and magnitude of the anisotropy level $\(\frac{\Delta \omm}{\bomm}\)^{U2}$. The view point is on top of the hemisphere of maximum (left) and minimum (right) acceleration.

 In an effort to identify possible redshift dependence of the above anisotropy, we have implemented a redshift tomography of the Union2 data and have identified the maximum anisotropy directions (with their errors) for the following redshift ranges:
0-0.2, 0-0.4, 0-0.6, 0-0.8, 0-1.0, 0-1.2. 0-1.4 (see Table 1 and Fig. 3). We have found that with the exception of the 0-0.2 redshift range whose maximum anisotropy direction  is about $40^\circ$ away from the maximum anisotropy direction of the full dataset, all the other redshift ranges have an anisotropy direction which is within about $20^\circ$ from the anisotropy direction of the full Union2 dataset.

\vspace{0pt}
\begin{table*}[t!]
\begin{center}
\caption{Directions of maximum anisotropy for several redshift ranges of the Union2 data (see also Fig. 3). The level of maximum anisotropy for a typical isotropic simulated dataset is also shown in the $5^{th}$ column. The asymmetry of errors is largely due to the non-uniform distribution of the SnIa on the sky.}
\begin{tabular}{ccccc}
\hline
\hline\\
\vspace{3pt}\textbf{Redshift Range} \hspace{3pt}& \textbf{\it l}\hspace{3pt} & \textbf{\it b}\hspace{3pt}& \textbf{\it $\(\frac{\Delta \ommax}{\bomm}\)^{U2}$}\hspace{6pt}& \textbf{\it $\(\frac{\Delta \ommax}{\bomm}\)^{SIM}$}\hspace{6pt}  \\
\vspace{3pt} 0-0.2                & ${341^\circ}^{+9}_{-22}$               \hspace{3pt} &  ${-17^\circ}^{+28}_{-6}$      \hspace{3pt} &  $2.08\pm 0.22$              &  $4.28\pm 0.22$ \\
\vspace{3pt} 0-0.4                   & ${301^\circ}^{+16}_{-2}$               \hspace{3pt} &  ${-1^\circ}^{+21}_{-15}$       \hspace{3pt} &   $1.81\pm 0.27$  &  $1.23\pm 0.15$        \\
\vspace{3pt} 0-0.6                & ${301^\circ}^{+43}_{-14}$              \hspace{3pt} &  ${-4^\circ}^{+23}_{-26}$      \hspace{3pt} & $0.6\pm 0.1$    &  $0.48\pm 0.08$               \\
\vspace{3pt} 0-0.8          & ${327^\circ}^{+22}_{-21}$               \hspace{3pt} &  ${37^\circ}^{+4}_{-19}$   \hspace{3pt} &  $0.46\pm 0.07$    &  $0.36\pm 0.06$               \\
\vspace{3pt} 0-1.0             & ${301^\circ}^{+35}_{-0}$              \hspace{3pt} &  ${-4^\circ}^{+31}_{-0}$   \hspace{3pt} &  $0.45\pm 0.07$   &  $0.35\pm 0.06$              \\
\vspace{3pt} 0-1.2                & ${310^\circ}^{+8}_{-4}$              \hspace{3pt} &  ${16^\circ}^{+16}_{-11}$      \hspace{3pt} &  $0.43\pm 0.07$    &  $0.35\pm 0.06$                \\
\hline \\
\vspace{3pt} 0-1.4 (all data)               & ${309^\circ}^{+23}_{-3}$               \hspace{3pt} & ${18^\circ}^{+11}_{-10}$    \hspace{3pt} &           $0.43\pm 0.06$      &  $0.36\pm 0.06$            \\
\hline \hline
\end{tabular}
\end{center}
\end{table*}

We next wish to address the question whether the maximum anisotropy level (\ref{maxlevel}) for the Union2 data is consistent with statistical isotropy. In order to address this question we have constructed simulated isotropic datasets by replacing the $i^{th}$ distance modulus of the Union2 dataset by a random number with a gaussian distribution with mean and standard deviation determined by the best fit value of $\mu_{th}(z_i,\omm,\mu_0)$ (eq. (\ref{mth}) with the best fit values $\omm=0.27$, $\mu_0=43.16$) and by $\sigma_{\mu\; i}$ of the corresponding Union2 data-point respectively. We then compare a simulated isotropic dataset with the real Union2 dataset by spliting each dataset into hemisphere pairs using 10 random directions. This part of our analysis is not aimed at identifying the maximum anisotropy direction neither at comparing with the result of the search in N=400 directions. Instead it only aims at comparing the real data with the isotropic simulated data with respect to the level of anisotropy. In the context of this comparison it is not important to identify the level of absolute maximum anisotropy. What is more important is to repeat the comparison a relatively large number of times (40 in our case) in order to have acceptable statistics. Given the limitations of computing time we had to reduce the number of axes directions (10) in order to increase the number of Monte Carlo experiments performed. Clearly the level of anisotropy identified in this case ($\frac{\Delta \Omega_m}{\Omega_m}\simeq 0.29$) is significantly smaller compared to the case of 400 axes directions ($\frac{\Delta \Omega_m}{\Omega_m}\simeq 0.43$) but this is not important for our purposes which do not include in this case the identification of the maximum anisotropy. Next we find the maximum levels of anisotropy $\(\frac{\Delta \ommax}{\bomm}\)^{U2}$ (Union2) and $\(\frac{\Delta \ommax}{\bomm}\)^{I}$ (Isotropic) and compare them. We repeat this comparison experiment 40 times with different simulated isotropic data and axes each time. We found the following results:
\begin{figure}[!b]
\begin{center}
\includegraphics[width=3.5in]{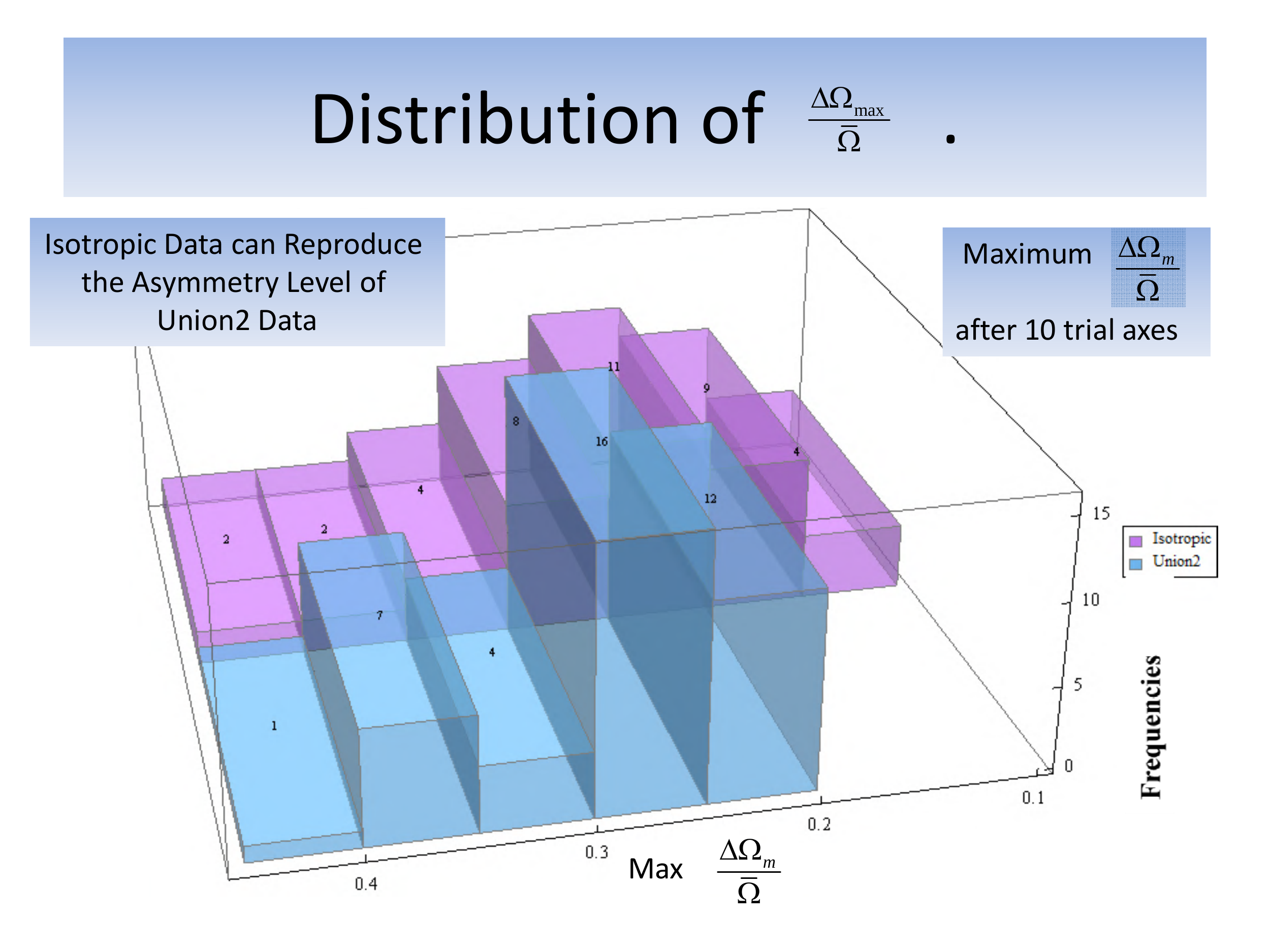}
\end{center}
\caption{\small The distribution of the anisotropy levels $\(\frac{\Delta \ommax}{\bomm}\)$ in the Union2 (front histogram) and simulated Isotropic datasets (back) using 10 random directions in each dataset. They show significant overlap which is a sign of consistency of the Union2 data with statistical isotropy.}
  \label{fig3}
\end{figure}
\begin{itemize}
\item
In about $1/3$ of the numerical experiments ($14$ times out of $40$) $\(\frac{\Delta \ommax}{\bomm}\)^{I}>\(\frac{\Delta \ommax}{\bomm}\)^{U2}$ \ie the anisotropy level was larger in the isotropic simulated data. In the rest $2/3$ of the numerical experiments the anisotropy level was larger in the Union2 data. This is a clear indication that the anisotropy level found in the Union2 data is consistent with statistical isotropy.
\item
The mean and standard deviation of the maximum anisotropy levels in each case are
\be
\(\frac{\Delta \ommax}{\bomm}\)^{U2}=0.29 \pm 0.05 \label{msdunau} \ee
for the Union2 data and
\be
\(\frac{\Delta \ommax}{\bomm}\)^{I}=0.24 \pm 0.07 \label{msdunai} \ee
for the simulated isotropic data.
 The error region described by equation (\ref{msdunau}) is not the error associated
with the uncertainties of the supernova magnitudes as are the error regions of the $4^{th}$ and $5^{th}$ columns of Table 1. It corresponds to the range of the
anisotropy level obtained when using 10 test axes to find the
maximum anisotropy. Given the relatively small number of axes
directions (10) considered in this part of the analysis, there
is a significant variation of the level of maximum anisotropy
identified in each run. This variation is described by the
result of eq. (\ref{msdunau}). In this case what we call `the error'
corresponds to the range of  values $\frac{\Delta
\Omega_{0m,max}}{\Omega_m}$ that is expected to be obtained in
about $68\%$ of the trials when our approach is implemented
using 10 axes with random directions. In view of the fact that
this variation is large enough to overlap significantly with
the corresponding results obtained with the isotropic Monte
Carlo data, it becomes clear that the anisotropy of the real
data is consistent with statistical isotropy.
It is clear from eqs. (\ref{msdunau}) and (\ref{msdunai}) that there is a clear overlap at the $1\sigma$ level which also implies that the maximum anisotropy level of the Union2 data is consistent with statistical isotropy.
\item
The histograms indicating the distribution of $\(\frac{\Delta \ommax}{\bomm}\)$ in each case are shown in Fig. 3 and they clearly show a significant overlap confirming also the consistency of the Union2 data with statistical isotropy.
\end{itemize}
Thus we have identified a direction of maximum anisotropy in the Union2 data and the level of this anisotropy is larger than about $70\%$ of isotropic simulated datasets. However this level is clearly not enough to indicate inconsistency with statistical isotropy.

In the next section we proceed to compare the direction of the identified maximum anisotropy with corresponding directions obtained with different, independent cosmological observations. The potential consistency among these independent anisotropy directions can dramatically increase the statistical significance of each one of them.

\section{Correlations with Preferred Axes from Other Observations}

\begin{figure}[!b]
\begin{center}
\includegraphics[width=3.5in]{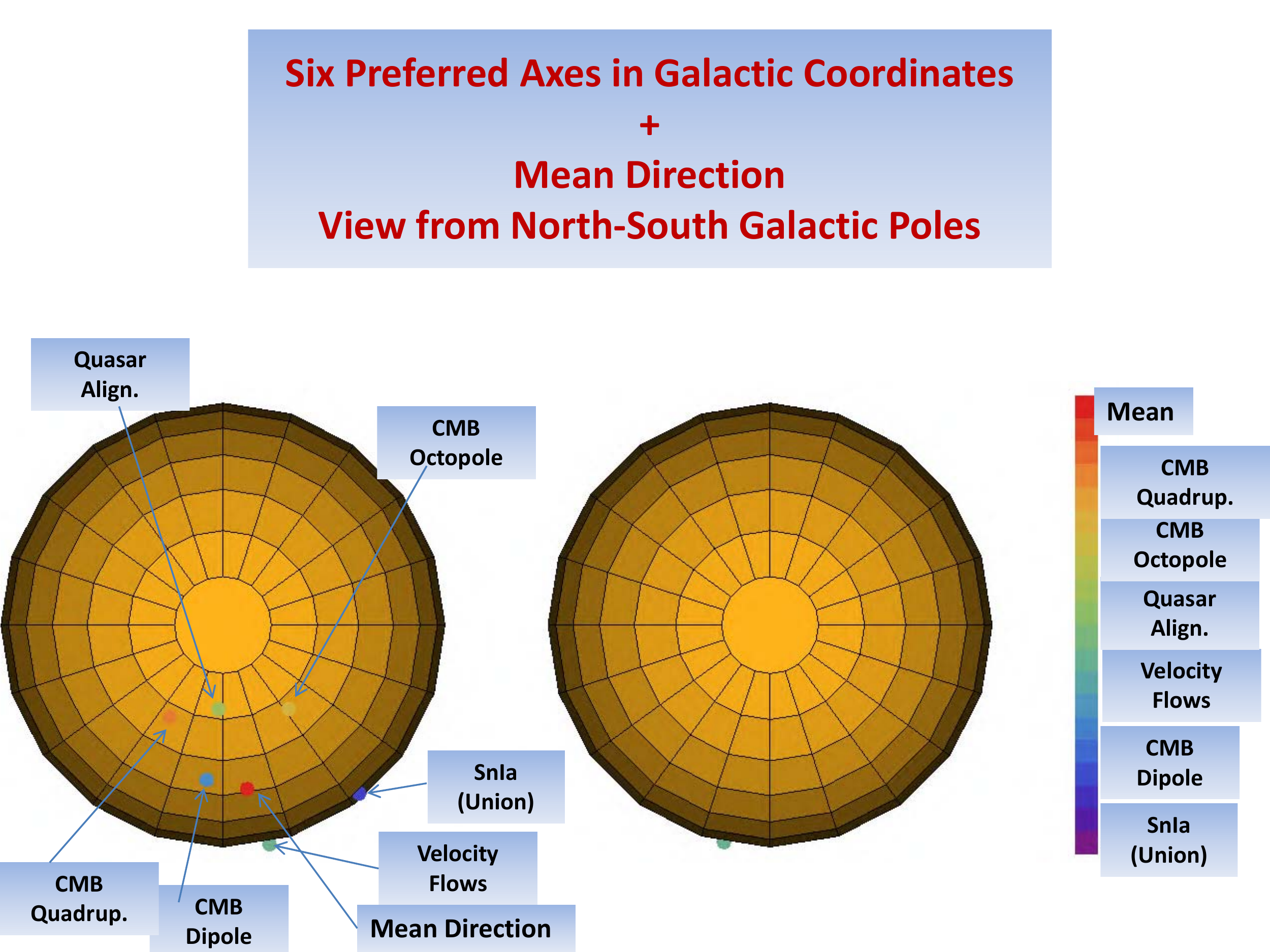}
\end{center}
\caption{\small The coordinates of the preferred axes of Table 2 are all located in a region less than a quarter of the North Galactic Hemisphere (left). The south galactic hemisphere (right) is also shown for completeness. The bulk flow direction is also visible in the south galactic hemisphere because it is close to the equator. The mean direction obtained in Table 2 with coordinates $(l,b)=(278^\circ,44^\circ)$ is also shown.}
  \label{fig4}
\end{figure}

As discussed in the Introduction, there is a range of independent cosmological observations which indicate the existence of anisotropy axes. The consistency of these directions may be interpreted as a hint of the existence of an underlying physical or systematic cause which is common in all of these apparently independent axes. These cosmological observations along with their preferred directions and the corresponding references are summarized in Table 2. In what follows we focus on the preferred {\it axes} and ignore the information about the directionality of each axis. Had we considered also information about the directionality of axes the likelihood of the observed coincidence would be even smaller given that the directions of the bulk velocity flow, faster accelerating expansion and CMB dipole all appear to be towards the North Galactic Hemisphere.

\vspace{0pt}
\begin{table}[!t]
\begin{center}
\caption{Directions of Preferred axes from different cosmological observations.The precise directions may vary by a few degrees across the literature but our results are insensitive to such small variations.}
\begin{tabular}{cccc}
\hline
\hline\\
\vspace{3pt}\textbf{Cosmological Obs.} \hspace{0pt}& \textbf{\it l}\hspace{3pt} & \textbf{\it b}\hspace{3pt}
&\textbf{Reference}\hspace{1pt}  \\
\vspace{3pt} SnIa Union2                  & $309^\circ$               \hspace{7pt} &  $18^\circ$      \hspace{7pt} &  This Study              \\
\vspace{3pt} CMB Dipole                     & $264^\circ$                \hspace{7pt} &  $48^\circ$       \hspace{7pt} &   \cite{Lineweaver:1996xa}          \\
\vspace{3pt} Velocity Flows                  & $282^\circ$               \hspace{7pt} &  $6^\circ$      \hspace{7pt} &  \cite{Feldman:2009es} \cite{Watkins:2008hf}                  \\
\vspace{3pt} Quasar Alignment           & $267^\circ$                \hspace{7pt} &  $69^\circ$   \hspace{7pt} &  \cite{Hutsemekers:2005iz}                    \\
\vspace{3pt} CMB Octopole             & $308^\circ$                \hspace{7pt} &  $63^\circ$   \hspace{7pt} &  \cite{Bielewicz:2004en}                  \\
\vspace{3pt} CMB Quadrupole                & $240^\circ$               \hspace{7pt} &  $63^\circ$      \hspace{7pt} &  \cite{Frommert:2009qw}  \cite{Bielewicz:2004en}                  \\
\hline \\
\vspace{3pt} Mean               & $278^\circ \pm 26^\circ$               \hspace{7pt} & $45^\circ \pm 27^\circ$    \hspace{7pt} &           -                  \\
\hline \hline
\end{tabular}
\end{center}
\end{table}

The six axes corresponding to the observations of Table 2 are shown in Fig. 4 in galactic coordinates. Clearly all coordinates are located in a relatively small part of the North Galactic Hemisphere (less than a quarter of it). It is straightforward to estimate the probability that six random points would lie is such a small region on a hemisphere. To obtain this estimate we evaluate the mean value of the inner product between all pairs of unit vectors corresponding to the preferred directions of Table 2 and compare with the corresponding mean value of six random directions on a hemisphere.
\begin{figure}[!b]
\begin{center}
\includegraphics[width=3.5in]{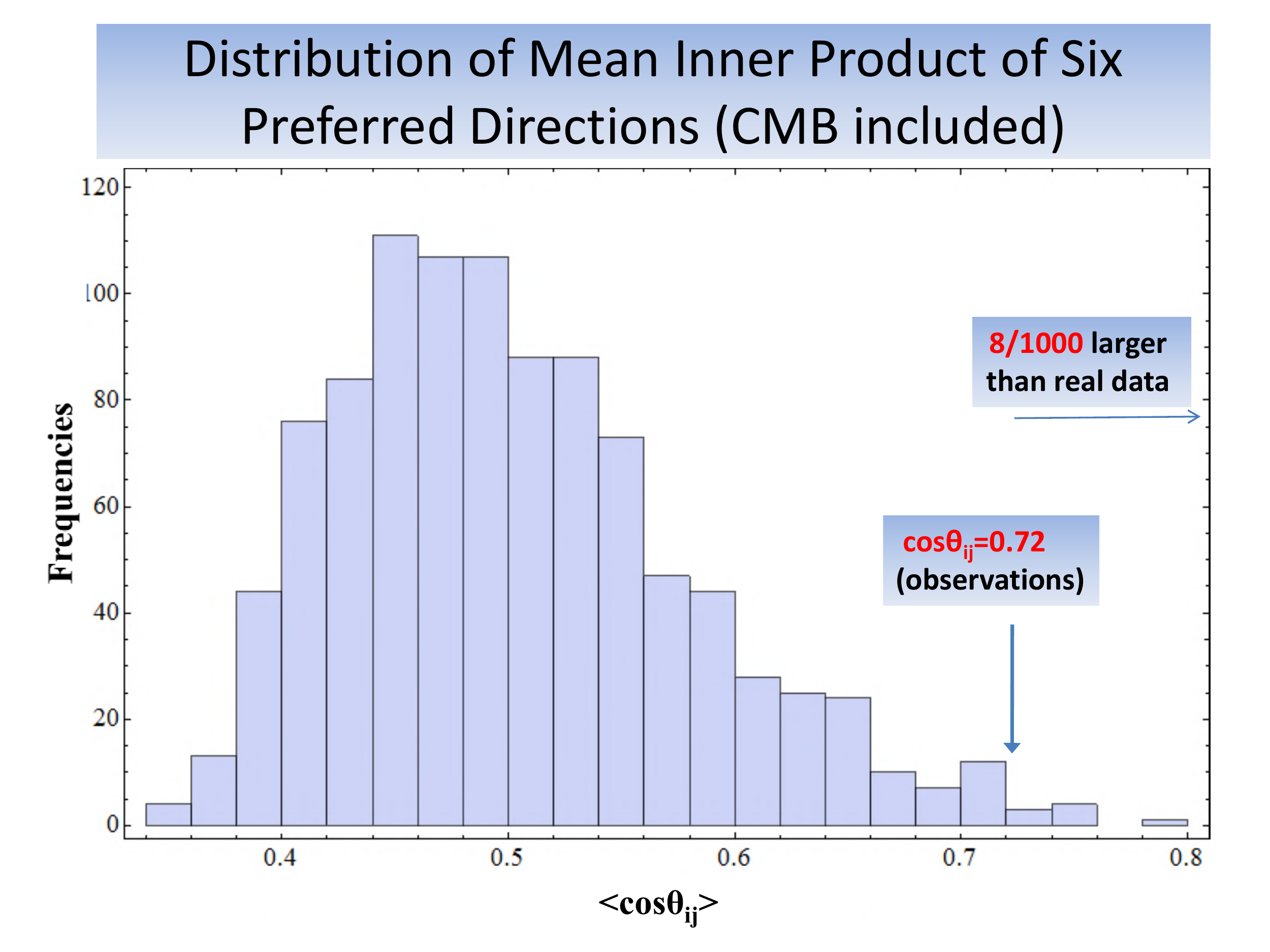}
\end{center}
\caption{\small A histogram of the distribution of $<\vert cos\theta_{ij}\vert>$ as obtained from the Monte Carlo data, superposed with the value obtained from the real data of Table 2 (eq. (\ref{mcsijrdu})). Less than $1\%$ of the Monte Carlo data exceed the value corresponding to the real data.}
  \label{fig5}
\end{figure}
Thus we evaluate
\be
<\vert cos\theta_{ij}\vert>=<\vert{\hat r}_i \cdot {\hat r}_j\vert>=\sum_{i,j=1, j\neq i}^N\frac{\vert{\hat r}_i \cdot {\hat r}_j\vert}{N(N-1)} \label{mcosij} \ee where in our case $N=6$ and we take the absolute value because we are ignoring the directionality of the axes. We thus apply eq. (\ref{mcosij}) to both the real data of Table 2 and to 1000 realizations of six random points on the sphere obtained using eq. (\ref{randdir}). For the real data we find \be <\vert cos\theta_{ij}\vert>=0.72 \label{mcsijrdu} \ee while from the Monte Carlo uncorrelated data we obtain \be <\vert cos\theta_{ij}\vert>=0.5\pm 0.072 \label{mcsijrdi} \ee Clearly, the value of $<\vert cos\theta_{ij}\vert>$ of the real data is about $3\sigma$ away from the expected value if there were no correlation among the axes of Table 2. This is also seen in Fig. 5 which shows a histogram of the distribution of $<\vert cos\theta_{ij}\vert>$ as obtained from the Monte Carlo data superposed with the value obtained from the real data (eq. (\ref{mcsijrdu}). Clearly, less than $1\%$ ($0.8\%$) of the simulated data exceed the value of $<\vert cos\theta_{ij}\vert>$ obtained with the real data. Thus, the coincidence of these independent preferred axes in such a small angular region is a highly unlikely event.
\begin{figure}[!b]
\begin{center}
\includegraphics[width=3.5in]{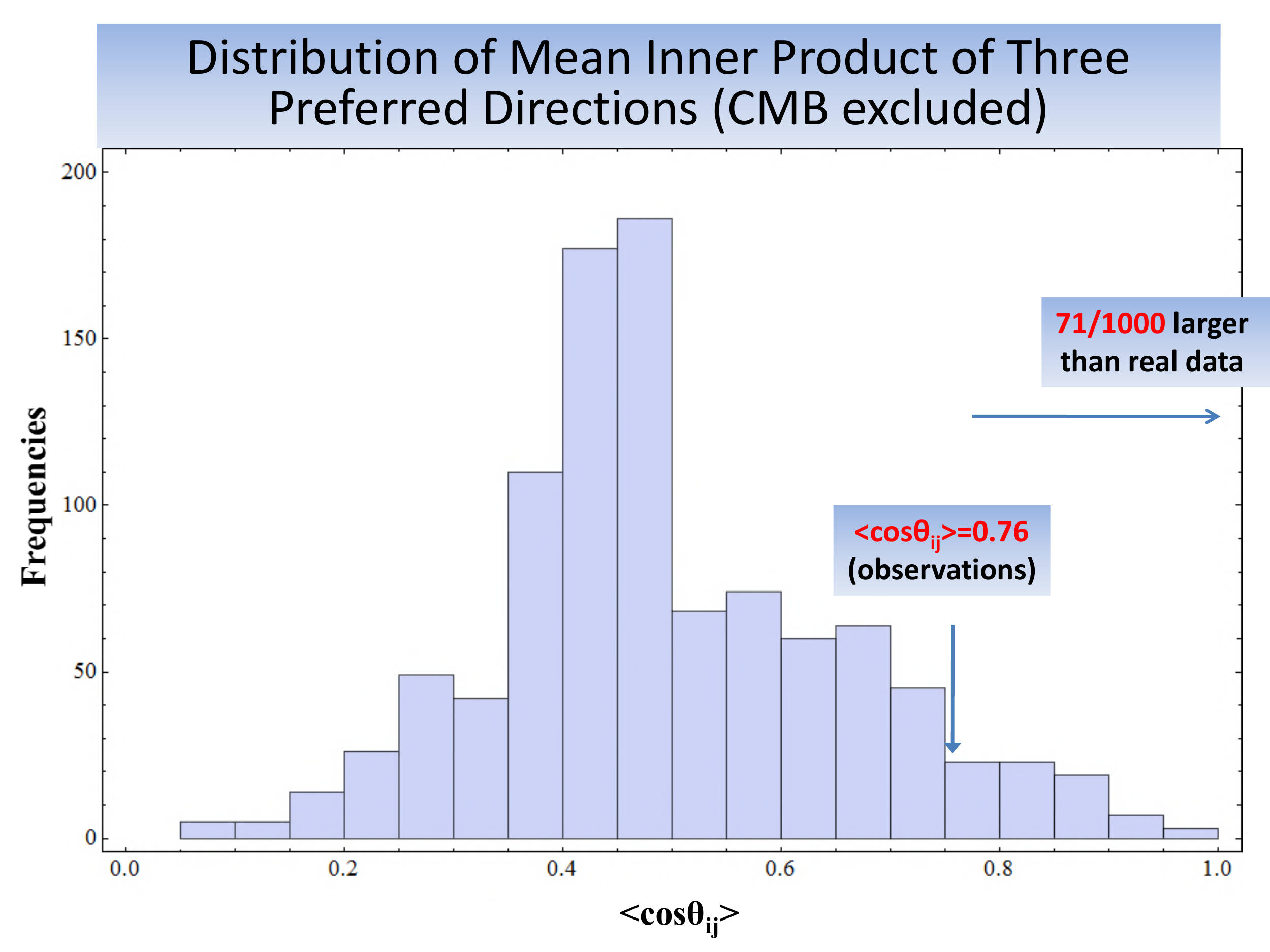}
\end{center}
\caption{\small A histogram of the distribution of $<\vert cos\theta_{ij}\vert>$ as obtained from the Monte Carlo data, superposed with the value obtained from the real data of Table 2 excluding the axes related to the CMB (eq. (\ref{mcsijrd3u})). About $7\%$ of the Monte Carlo data exceed the value corresponding to the real data.}
  \label{fig6}
\end{figure}
Even if we ignore the axes related to the CMB the coincidence of the three remaining axes in such a small angular region is still a relatively unlikely event with probability about $7\%$ (Fig. 6). In fact for the three remaining axes of Table 2 we find \be <\vert cos\theta_{ij}\vert>=0.76 \label{mcsijrd3u} \ee while the Monte Carlo simulation gives \be <\vert cos\theta_{ij}\vert>=0.5\pm 0.16 \label{mcsijrd3i} \ee \ie the real data are about $1.5\sigma$ away from the Monte Carlo mean value.

We therefore conclude that even though each of the axes of Table 2 does not by itself constitute statistically significant evidence for a cosmological anisotropy, their coexistence in a relatively small angular region is a very unlikely event which is most probably attributed to either an undiscovered physical effect or to a common basic systematic error that has so far escaped attention.

\section{Conclusions}
The main conclusions of our study may be summarized as follows:
\begin{itemize}
\item
The hemisphere of maximum accelerating expansion of the universe according to the Union2 data has a pole in the direction $(l,b)=({309^\circ}^{+23}_{-3}, {18^\circ}^{+11}_{-10})$ while the hemisphere of minimum acceleration is in the opposite direction.
\item
The level of anisotropy of the Union2 dataset is larger than about $70\%$ of simulated datasets but it is still consistent with statistical isotropy.
\item
The coincidence of the anisotropy axes of Table 2 in a relatively small angular region is statistically a highly unlikely event that hints towards a physical or systematic connection among the axes of Table 2.
\end{itemize}
 The combination of these cosmological observations may involve some a posteriori reasoning in the sense that there are other observations that show no hint of a preferred axis. However, most large scale cosmological observations may be analyzed in the context of a monopole, a dipole term (and higher moments) even if the significance level of a preferred direction is low and consistent with statistical isotropy. A real a posteriori reasoning would select {\it some} of these observations and points out that they point in similar directions. On the other hand our analysis includes {\it all} large scale cosmological observations that we could find in the literature which involve a preferred cosmological direction (even if it is of low significance by itself) and point out that these directions are abnormally close to each other.

The confirmation of the existence of a cosmological preferred axis would constitute a breakthrough in cosmological research. Given the present status of cosmological observations such a confirmation is one of the most probable directions from which new physics may emerge.

Given the preliminary evidence for anisotropy discussed in the present study, it is important to extend and intensify efforts for the possible confirmation of this evidence. Such confirmation may be achieved by extending the SnIa compilations towards larger datasets and deeper redshifts that span as uniformly as possible all directions in the sky. This is important in view of the fact that the Union2 compilation is less uniform and detailed in the south galactic hemisphere. In addition it is important to extend other cosmological data related to CMB low multipole moments, bulk velocity flows and quasar polarization to confirm the present existing evidence for preferred axes in these datasets. Finally, alternative probes of cosmological anisotropies may be considered like higher CMB multipole moments, non-gaussian features and polarization in the CMB maps, alignments of geometric features of various structures on large scales (there is already some preliminary evidence for alignment of handedness of spiral galaxies \cite{Longo:2009pr} along an axis not far from the directions of the other preferred axes of Table 2), alignment of optical polarization from various cosmological sources or studies based on cosmic parallax \cite{Quartin:2009xr}. It is also important to derive observational signatures that can clearly distinguish between the various different origins of the preferred axes discussed in the introduction.

{\bf Numerical Analysis Files:} The Mathematica 7 files with the data used for the production of the figures along with a Powerpoint file with additional figures may be found at http://leandros.physics.uoi.gr/anisotropy .

\section*{Acknowledgements}
We thank J.B. Sanchez for useful discussions. This work was supported by the European Research and Training
Network MRTN-CT-2006 035863-1 (UniverseNet).


\begin{thebibliography}{99}
\bibitem{Komatsu:2010fb}
  E.~Komatsu {\it et al.},
  arXiv:1001.4538 [astro-ph.CO].

\bibitem{Tegmark:2003ve}
  M.~Tegmark, A.~de Oliveira-Costa and A.~Hamilton,
  Phys.\ Rev.\  D {\bf 68}, 123523 (2003)
  [arXiv:astro-ph/0302496].

\bibitem{Copi:2010na}
  C.~J.~Copi, D.~Huterer, D.~J.~Schwarz and G.~D.~Starkman,
  arXiv:1004.5602 [astro-ph.CO].

\bibitem{Bennett:2010jb}
  C.~L.~Bennett {\it et al.},
  arXiv:1001.4758 [Unknown].

\bibitem{Schwarz:2004gk}
  D.~J.~Schwarz, G.~D.~Starkman, D.~Huterer and C.~J.~Copi,
  Phys.\ Rev.\ Lett.\  {\bf 93}, 221301 (2004)
  [arXiv:astro-ph/0403353].

\bibitem{Land:2005ad}
  K.~Land and J.~Magueijo,
  Phys.\ Rev.\ Lett.\  {\bf 95}, 071301 (2005)
  [arXiv:astro-ph/0502237].

\bibitem{Gruppuso:2010up}
  A.~Gruppuso and K.~M.~Gorski,
  JCAP {\bf 1003}, 019 (2010)
  [arXiv:1002.3928 [Unknown]].

\bibitem{Sarkar:2010yj}
  D.~Sarkar, D.~Huterer, C.~J.~Copi, G.~D.~Starkman and D.~J.~Schwarz,
  arXiv:1004.3784 [Unknown].

\bibitem{Hanson:2009gu}
  D.~Hanson and A.~Lewis,
  Phys.\ Rev.\  D {\bf 80}, 063004 (2009)
  [arXiv:0908.0963 [astro-ph.CO]].

\bibitem{Smith:2009jr}
  K.~M.~Smith, L.~Senatore and M.~Zaldarriaga,
  JCAP {\bf 0909}, 006 (2009)
  [arXiv:0901.2572 [astro-ph]].

\bibitem{Amanullah:2010vv}
  R.~Amanullah {\it et al.},
  Astrophys.\ J.\  {\bf 716}, 712 (2010)
  [arXiv:1004.1711 [Unknown]].

\bibitem{lpsniarev} J. C. Bueno Sanchez, S. Nesseris and L. Perivolaropoulos
JCAP {\bf 0911}, 029 (2009)
[arXiv:0908.2636 [astro-ph]].

\bibitem{Nesseris:2007pa}
  S.~Nesseris and L.~Perivolaropoulos,
  Phys.\ Rev.\  D {\bf 77}, 023504 (2008)
  [arXiv:0710.1092 [astro-ph]].

\bibitem{Perivolaropoulos:2008ud}
  L.~Perivolaropoulos,
  arXiv:0811.4684 [astro-ph]; R.~J.~Yang and S.~N.~Zhang,
  arXiv:0905.2683 [astro-ph.CO];

\bibitem{TrujilloGomez:2010yh}
  S.~Trujillo-Gomez, A.~Klypin, J.~Primack and A.~J.~Romanowsky,
  arXiv:1005.1289 [astro-ph.CO].

\bibitem{Reid:2009xm}
  B.~A.~Reid {\it et al.},
  Mon.\ Not.\ Roy.\ Astron.\ Soc.\  {\bf 404}, 60 (2010)
  [arXiv:0907.1659 [astro-ph.CO]].

\bibitem{Watkins:2008hf}
  R.~Watkins, H.~A.~Feldman and M.~J.~Hudson,
  arXiv:0809.4041 [astro-ph].

\bibitem{Lavaux:2008th}
  G.~Lavaux, R.~B.~Tully, R.~Mohayaee and S.~Colombi,
  Astrophys.\ J.\  {\bf 709}, 483 (2010)
  [arXiv:0810.3658 [astro-ph]].

\bibitem{Kashlinsky:2008ut}
  A.~Kashlinsky, F.~Atrio-Barandela, D.~Kocevski and H.~Ebeling,
  arXiv:0809.3734 [astro-ph].

\bibitem{Tsagas:2009nh}
  C.~G.~Tsagas,
  Mon.\ Not.\ Roy.\ Astron.\ Soc.\  {\bf 405}, 503 (2010)
  [arXiv:0902.3232 [astro-ph.CO]].

\bibitem{Bielewicz:2004en}
  P.~Bielewicz, K.~M.~Gorski and A.~J.~Banday,
  Mon.\ Not.\ Roy.\ Astron.\ Soc.\  {\bf 355}, 1283 (2004)
  [arXiv:astro-ph/0405007].

\bibitem{Frommert:2009qw}
  M.~Frommert and T.~A.~Ensslin,
  arXiv:0908.0453 [astro-ph.CO].

\bibitem{Lineweaver:1996xa}
  C.~H.~Lineweaver, L.~Tenorio, G.~F.~Smoot, P.~Keegstra, A.~J.~Banday and P.~Lubin,
  Astrophys.\ J.\  {\bf 470}, 38 (1996)
  [arXiv:astro-ph/9601151].

\bibitem{Kovetz:2010kv}
  E.~D.~Kovetz, A.~Ben-David and N.~Itzhaki,
  arXiv:1005.3923 [astro-ph.CO].

\bibitem{Hutsemekers:2005iz}
  D.~Hutsemekers, R.~Cabanac, H.~Lamy and D.~Sluse,
  Astron.\ Astrophys.\  {\bf 441}, 915 (2005)
  [arXiv:astro-ph/0507274];
  D.~Hutsemekers, A.~Payez, R.~Cabanac, H.~Lamy, D.~Sluse, B.~Borguet and J.~R.~Cudell,
  arXiv:0809.3088 [astro-ph];
  D.~Hutsemekers and H.~Lamy,
  arXiv:astro-ph/0012182.

\bibitem{Broadhurst:2004bi}
  T.~J.~Broadhurst, M.~Takada, K.~Umetsu, X.~Kong, N.~Arimoto, M.~Chiba and T.~Futamase,
  Astrophys.\ J.\  {\bf 619}, L143 (2005)
  [arXiv:astro-ph/0412192].

\bibitem{Umetsu:2007pq}
  K.~Umetsu and T.~Broadhurst,
  Astrophys.\ J.\  {\bf 684}, 177 (2008)
  [arXiv:0712.3441 [astro-ph]].

\bibitem{Copi:2006tu}
  C.~Copi, D.~Huterer, D.~Schwarz and G.~Starkman,
  Phys.\ Rev.\  D {\bf 75}, 023507 (2007)
  [arXiv:astro-ph/0605135].

\bibitem{Klypin:1999uc}
  A.~A.~Klypin, A.~V.~Kravtsov, O.~Valenzuela and F.~Prada,
  Astrophys.\ J.\  {\bf 522}, 82 (1999)
  [arXiv:astro-ph/9901240].

\bibitem{Moore:2001fc}
  B.~Moore,
  arXiv:astro-ph/0103100.

\bibitem{Madau:2008fr}
  P.~Madau, J.~Diemand and M.~Kuhlen,
  arXiv:0802.2265 [astro-ph].

\bibitem{cuspygal}G. Gentile et. al.
MNRAS, {\bf 351}, 903 (2004); G. Gentile et. al.
ApJ, {\bf 634}, L145 (2005);  J. D. Simon et. al., ApJ, {\bf 621}, 757 (2005); W. J. de Blok, ApJ, {\bf 634}, 227 (2005).

\bibitem{Peiris:2010jd}
  H.~V.~Peiris and T.~L.~Smith,
  Phys.\ Rev.\  D {\bf 81}, 123517 (2010)
  [arXiv:1002.0836 [Unknown]].

\bibitem{Zumalacarregui:2010wj}
  M.~Zumalacarregui, T.~S.~Koivisto, D.~F.~Mota and P.~Ruiz-Lapuente,
  JCAP {\bf 1005}, 038 (2010)
  [arXiv:1004.2684 [astro-ph.CO]].

\bibitem{Koivisto:2005mm}
  T.~Koivisto and D.~F.~Mota,
  Phys.\ Rev.\  D {\bf 73}, 083502 (2006)
  [arXiv:astro-ph/0512135].

\bibitem{Battye:2009ze}
  R.~Battye and A.~Moss,
  Phys.\ Rev.\  D {\bf 80}, 023531 (2009)
  [arXiv:0905.3403 [astro-ph.CO]].

\bibitem{ArmendarizPicon:2004pm}
  C.~Armendariz-Picon,
  JCAP {\bf 0407}, 007 (2004)
  [arXiv:astro-ph/0405267].

\bibitem{EspositoFarese:2009aj}
  G.~Esposito-Farese, C.~Pitrou and J.~P.~Uzan,
  Phys.\ Rev.\  D {\bf 81}, 063519 (2010)
  [arXiv:0912.0481 [Unknown]].

\bibitem{Rodrigues:2007ny}
  D.~C.~Rodrigues,
  Phys.\ Rev.\  D {\bf 77}, 023534 (2008)
  [arXiv:0708.1168 [astro-ph]].

\bibitem{Jimenez:2008vs}
  J.~B.~Jimenez and A.~L.~Maroto,
  JCAP {\bf 0903}, 015 (2009)
  [arXiv:0811.3606 [astro-ph]].

\bibitem{Alexander:2007xx}
  S.~Alexander, T.~Biswas, A.~Notari and D.~Vaid,
  JCAP {\bf 0909}, 025 (2009)
  [arXiv:0712.0370 [astro-ph]].

\bibitem{GarciaBellido:2008nz}
  J.~Garcia-Bellido and T.~Haugboelle,
  JCAP {\bf 0804}, 003 (2008)
  [arXiv:0802.1523 [astro-ph]].

\bibitem{Biswas:2010xm}
  T.~Biswas, A.~Notari and W.~Valkenburg,
  arXiv:1007.3065 [Unknown].

\bibitem{Dunsby:2010ts}
  P.~Dunsby, N.~Goheer, B.~Osano and J.~P.~Uzan,
  JCAP {\bf 1006}, 017 (2010)
  [arXiv:1002.2397 [Unknown]].

\bibitem{Garfinkle:2009uf}
  D.~Garfinkle,
  Class.\ Quant.\ Grav.\  {\bf 27} (2010) 065002
  [arXiv:0908.4102 [gr-qc]].

\bibitem{Luminet:2008ew}
  J.~P.~Luminet,
  arXiv:0802.2236 [astro-ph].

\bibitem{Bielewicz:2008ga}
  P.~Bielewicz and A.~Riazuelo,
  arXiv:0804.2437 [astro-ph].

\bibitem{Carneiro:2001fz}
  S.~Carneiro and G.~A.~Mena Marugan,
  Phys.\ Rev.\  D {\bf 64}, 083502 (2001)
  [arXiv:gr-qc/0109039].

\bibitem{Akofor:2007fv}
  E.~Akofor, A.~P.~Balachandran, S.~G.~Jo, A.~Joseph and B.~A.~Qureshi,
  JHEP {\bf 0805}, 092 (2008)
  [arXiv:0710.5897 [astro-ph]].

\bibitem{Koivisto:2010dr}
  T.~S.~Koivisto, D.~F.~Mota, M.~Quartin and T.~G.~Zlosnik,
  arXiv:1006.3321 [Unknown].

\bibitem{ArmendarizPicon:2007nr}
  C.~Armendariz-Picon,
  JCAP {\bf 0709}, 014 (2007)
  [arXiv:0705.1167 [astro-ph]].

\bibitem{Pullen:2007tu}
  A.~R.~Pullen and M.~Kamionkowski,
  Phys.\ Rev.\  D {\bf 76}, 103529 (2007)
  [arXiv:0709.1144 [astro-ph]].

\bibitem{Ackerman:2007nb}
  L.~Ackerman, S.~M.~Carroll and M.~B.~Wise,
  Phys.\ Rev.\  D {\bf 75}, 083502 (2007)
  [Erratum-ibid.\  D {\bf 80}, 069901 (2009)]
  [arXiv:astro-ph/0701357].

\bibitem{ValenzuelaToledo:2010cs}
  C.~A.~Valenzuela-Toledo,
  arXiv:1004.5363 [Unknown].

\bibitem{Dimopoulos:2008yv}
  K.~Dimopoulos, M.~Karciauskas, D.~H.~Lyth and Y.~Rodriguez,
  JCAP {\bf 0905}, 013 (2009)
  [arXiv:0809.1055 [astro-ph]].

\bibitem{Yokoyama:2008xw}
  S.~Yokoyama and J.~Soda,
  JCAP {\bf 0808}, 005 (2008)
  [arXiv:0805.4265 [astro-ph]].

\bibitem{Golovnev:2009ks}
  A.~Golovnev and V.~Vanchurin,
  Phys.\ Rev.\  D {\bf 79}, 103524 (2009)
  [arXiv:0903.2977 [astro-ph.CO]].

\bibitem{Bartolo:2009pa}
  N.~Bartolo, E.~Dimastrogiovanni, S.~Matarrese and A.~Riotto,
  JCAP {\bf 0910}, 015 (2009)
  [arXiv:0906.4944 [astro-ph.CO]].

\bibitem{ghosts}
  B.~Himmetoglu, C.~R.~Contaldi and M.~Peloso,
  Phys.\ Rev.\ Lett.\  {\bf 102}, 111301 (2009)
  [arXiv:0809.2779 [astro-ph]]; B.~Himmetoglu, C.~R.~Contaldi and M.~Peloso,
  Phys.\ Rev.\  D {\bf 79}, 063517 (2009)
  [arXiv:0812.1231 [astro-ph]]; B.~Himmetoglu, C.~R.~Contaldi and M.~Peloso,
  Phys.\ Rev.\  D {\bf 80}, 123530 (2009)
  [arXiv:0909.3524].


\bibitem{Kahniashvili:2008sh}
  T.~Kahniashvili, G.~Lavrelashvili and B.~Ratra,
  Phys.\ Rev.\  D {\bf 78}, 063012 (2008)
  [arXiv:0807.4239 [astro-ph]].

\bibitem{Barrow:1997mj}
  J.~D.~Barrow, P.~G.~Ferreira and J.~Silk,
  Phys.\ Rev.\ Lett.\  {\bf 78}, 3610 (1997)
  [arXiv:astro-ph/9701063].

\bibitem{Campanelli:2009tk}
  L.~Campanelli,
  Phys.\ Rev.\  D {\bf 80}, 063006 (2009)
  [arXiv:0907.3703 [astro-ph.CO]].

\bibitem{Kim:2009gi}
  J.~Kim and P.~Naselsky,
  JCAP {\bf 0907}, 041 (2009)
  [arXiv:0903.1930 [astro-ph.CO]].

\bibitem{Blomqvist:2009ps}
  M.~Blomqvist and E.~Mortsell,
  JCAP {\bf 1005}, 006 (2010)
  [arXiv:0909.4723 [Unknown]].

\bibitem{Blomqvist:2010ky}
  M.~Blomqvist, J.~Enander and E.~Mortsell,
  arXiv:1006.4638 [Unknown].

\bibitem{Tomita:2001gh}
  K.~Tomita,
  Prog.\ Theor.\ Phys.\  {\bf 106}, 929 (2001)
  [arXiv:astro-ph/0104141].

\bibitem{Cooke:2009ws}
  R.~Cooke and D.~Lynden-Bell,
  arXiv:0909.3861 [astro-ph.CO].

\bibitem{Kolatt:2000yg}
  T.~S.~Kolatt and O.~Lahav,
  Mon.\ Not.\ Roy.\ Astron.\ Soc.\  {\bf 323}, 859 (2001)
  [arXiv:astro-ph/0008041].

\bibitem{Gupta:2007pb}
  S.~Gupta, T.~D.~Saini and T.~Laskar,
  arXiv:astro-ph/0701683.

\bibitem{Schwarz:2007wf}
  D.~J.~Schwarz and B.~Weinhorst,
  arXiv:0706.0165 [astro-ph].

\bibitem{Riess:2004nr}
  A.~G.~Riess {\it et al.}  [Supernova Search Team Collaboration],
  Astrophys.\ J.\  {\bf 607}, 665 (2004)
  [arXiv:astro-ph/0402512].


\bibitem{Kowalski:2008ez}
  M.~Kowalski {\it et al.},
  Astrophys.\ J.\  {\bf 686}, 749 (2008)
  [arXiv:0804.4142 [astro-ph]].

\bibitem{Hicken:2009dk}
  M.~Hicken {\it et al.},
  Astrophys.\ J.\  {\bf 700}, 1097 (2009)
  [arXiv:0901.4804 [astro-ph.CO]].

\bibitem{Holtzman:2008zz}
  J.~A.~Holtzman {\it et al.},
  Astron.\ J.\  {\bf 136}, 2306 (2008).

\bibitem{astrocalc} P. Duffett-Smith, `Practical Astronomy with your Calculator' Cambridge University Press (1989).
Duffett-Smith Peter

\bibitem{Gorski:2004by}
  K.~M.~Gorski, E.~Hivon, A.~J.~Banday, B.~D.~Wandelt, F.~K.~Hansen, M.~Reinecke and M.~Bartelman,
  Astrophys.\ J.\  {\bf 622}, 759 (2005)
  [arXiv:astro-ph/0409513].

\bibitem{Feldman:2009es}
  H.~A.~Feldman, R.~Watkins and M.~J.~Hudson,
  Mon.\ Not.\ Roy.\ Astron.\ Soc.\  {\bf 392}, 756 (2010)
  [arXiv:0911.5516 [Unknown]].

\bibitem{Longo:2009pr}
  M.~J.~Longo,
  arXiv:0904.2529 [astro-ph.CO].

\bibitem{Quartin:2009xr}
  M.~Quartin and L.~Amendola,
  Phys.\ Rev.\  D {\bf 81}, 043522 (2010)
  [arXiv:0909.4954 [Unknown]].

\end{thebibliography}
\end{document}